\documentclass[twocolumn,showpacs,preprintnumbers,amsmath,amssymb]{revtex4}

\usepackage{graphicx,color}
\usepackage{dcolumn}
\usepackage{bm}
\def\mib#1{\mbox{\boldmath $#1$}}

\begin{document}


\title{
Fluid dynamics of dilatant fluid
}

\author{Hiizu Nakanishi}
\affiliation{
Department of Physics, Kyushu University 33, Fukuoka 812-8581, Japan}

\author{Shin-ichiro Nagahiro}
\affiliation{
Department of Mechanical Engineering, 
Sendai National College of Technology,
Natori, Miyagi 981-1239, Japan
}

\author{Namiko Mitarai}%
\affiliation{
Niels Bohr Institute, University of Copenhagen, Blegdamsvej 17, DK-2100,
Copenhagen \O, Denmark }%

\date{\today}

\begin{abstract}
Dense mixture of granules and liquid often shows a sever shear
thickening and is called a dilatant fluid.  We construct 
a fluid dynamics model for the dilatant fluid 
by introducing a phenomenological
state variable for a local state of dispersed particles.
With simple assumptions for an equation of the state variable, we
demonstrate that the model can describe basic features of the
dilatant fluid such as the stress-shear rate curve that represents
discontinuous severe shear thickening, hysteresis upon changing shear
rate, instantaneous hardening upon external impact.
Analysis of the model reveals that the shear thickening fluid shows an
instability in a shear flow
%
%
for some regime and exhibits {\it the shear thickening oscillation},
i.e. the oscillatory shear flow alternating between the thickened and
the relaxed states.
Results of numerical simulations are presented for one and
two-dimensional systems.
%
%
%
%
\end{abstract}



\maketitle
\section{Introduction}

One of the most common materials of the dilatant fluid is a dense
mixture of cornstarch and water, and it can be used to demonstrate a
number of counter-intuitive behaviors that the shear thickening medium
shows: sudden solidification upon externally applied stress, quick
re-fluidization after removal of the stress, formation of holes and
protrusions under strong vibration\cite{Swinney-2004,Ebata-2009}, etc.

These behaviors come from severe shear thickening
 and hysteresis,
%
that dense colloid or dense mixture of granules and liquid often show.
The shear viscosity increases almost discontinuously by orders of
magnitude at a certain critical shear rate\cite{Bonn-2008}, which makes
the fluid almost rigid against the sudden application of stress.
It is called a ``dilatant fluid'' by analogy with the
behavior of a granular medium\cite{Freundlich-1935}; 
when a granular medium is densely packed in a bag
that is flexible but non-stretchable, it cannot be deformed because the
volume is constant.  The granular medium must dilate upon deformation
due to the {\em principle of dilatancy} by Reynolds\cite{Reynolds-1885}.

There are several peculiar features in the shear thickening of the
dilatant fluid:
(i) the thickening is so severe and instantaneous
that it might be used even to make a body armor to stop a
bullet\cite{Wagner-2009},
(ii) the relaxation after removal of the external stress occurs within a
few seconds, that is quick but not as instantaneous as in the
thickening process,
(iii) 
the medium in the thickened state 
behaves like a rigid material allowing little 
elastic deformation as long as it is under stress,
(iv) the viscosity shows hysteresis upon changing the shear
rate\cite{Laun-1991},
(v) noisy fluctuations have been observed in the response to an external
shear stress in the thickening regime\cite{Laun-1991,Lootens-2005}.


Despite of the apparent analogy between the behaviors shown by these
media, it is not clear if the shear thickening of the dilatant fluid has
something to do with the property of dilatancy of granular media.
Originally, the shear thickening in colloid systems were regarded as a
result of the disorder transition of the layer and/or string structure
developed in the low shear rate
regime\cite{Hoffman-1972,Hoffman-1974,Hoffman-1998,Barnes-1989}.  The
dispersed particles align due the shear flow to give shear thinning in
low shear regime, but the turbulent motion in high shear regime destroys
this structure to give shear thickening. 
Such layer and/or string
structures have been observed in numerical
simulations\cite{Erpenbeck-1984} and experiments\cite{Chow-1995}, and in
some cases the shear thickening occurs when the structure is
broken\cite{Hoffman-1972}.  However, there are some other cases where
no significant structure change are observed upon discontinuous shear
thickening\cite{Laun-1992,Wagner-1996,Wagner-2002,Wagner-2005}.
Hydrocluster formation has been proposed as an alternative origin of the
shear thickening\cite{Brady-1985,Wagner-1996,MelroseBall-2004}.  Due to
hydrodynamic interaction among particles in the fluid, there is a
certain condition that clusters of particles grow and they can give
large viscosity.  Such a cluster structure of particles has been first
identified in numerical simulations\cite{Brady-1985}, then suggested by
SANS experiments\cite{Wagner-1996}. 
%
More direct observation has been made using fast confocal
microscopy\cite{Cheng-2011}.  
%
Jamming is another
possibility under active debate in recent years in connection with the
glass transition.  In dense granular system, the jamming can cause the
divergence of
viscosity\cite{Bonn-2008,Jaeger-2009,MelroseBall-2004-2,Ball-1997,
Cates-1998,Bertrand-2002,Lootens-2005}.
In connection with the dilatancy, Brown and Jaeger
studied the discontinuous shear thickening
%
%
and obtained somewhat empirical constitutive relations\cite{Brown-2010}.

There are no microscopic theories for the dilatant fluid yet in the
sense that the shear thickening is derived from the elementary
interactions among constituents of the medium, i.e. granules and fluid,
but there are a couple of semi-empirical theories: the soft-glassy
rheology (SGR) model\cite{Head-2001} and the schematic mode coupling
theory (MCT)\cite{Holmes-2005}.  The SGR model is the model based on the
stochastic dynamics with the activation energy that depends on the
stress. This model is extended to describe the shear thickening by
introducing the stress dependent effective temperature.  MCT, which
gives reasonable description of the glass transition, has been extended
schematically by introducing shear rate dependent integral kernel.  Both
of the theories are semi-empirical and have been demonstrated to show
the discontinuous shear thickening, but they have not been incorporated
in the fluid dynamics to study its flowing behavior of the medium.

Recently, the present authors constructed a fluid dynamic model for
the dilatant fluid by phenomenologically introducing an internal state
variable,
which determines the viscosity of the medium\cite{Nakanishi-2011}.  
The state variable itself is determined by the local stress\cite{com-1}.
The purpose of this paper is to present
detailed study on the flowing property of the medium represented by the
model.
We demonstrate that the model shows the discontinuous shear thickening
transition and the hysteresis upon changing the shear rate as has been
observed in experiments.  It is also shown that the steady shear flow
becomes unstable for a certain parameter range against the shear
thickening oscillation, where the medium alternates between the
thickened and the relaxed states.

The paper is organized as follow.  The model is introduced in Sec.2, and
it is examined for a simple uniform shear flow configuration in Sec.3.
Similar analysis is given for the gravitational slope flow and
Poiseuille flow in Sec.4.  The response to an impact is simulated in
Sec.5.  Effects of inhomogeneity is studied in Sec.6 by two-dimensional
simulations.  Summary and discussions are given in Sec.7.

\section{Model}

The model is based on the fluid dynamics with an internal state
variable that describes the local structure of particles dispersed in
the liquid.  The viscosity of the medium is determined by the internal
state, which in turn changes in response to the local shear stress.  
We introduce each element of the model in the following.

\paragraph{Fluid dynamics:}

The dynamics of the medium as a fluid is represented by the velocity
field $\mib v(\mib r)$, and is governed by the hydrodynamic
equation,
\begin{equation}
\rho{D v_i\over Dt} = 
{\partial\over\partial x_j}\Bigl( -P\,\delta_{i,j}+\sigma_{i,j}\Bigr)+\rho g_i,
\label{eq.v}
\end{equation}
where the Lagrange derivative is introduced:
\begin{equation}
{D\over Dt} \equiv 
    {\partial \over\partial t}+v_j{\partial\over\partial x_j} .
\end{equation}
The symbols $\rho$, $P$, and $\sigma_{i,j}$ represent the density, the
pressure, and the $(i, j)$ component of the viscous stress tensor
$\hat\sigma$, respectively.  The last term in Eq.(\ref{eq.v}) represents
the body force on the fluid due to the gravitational acceleration $g_i$.
We employ Einstein's rule for the summation over repeated suffixes.

We consider the incompressible fluid, thus the pressure $P$ is
determined by the incompressible condition
\begin{equation}
\mib\nabla\cdot\mib v(\mib r) = 0.
\label{incomp}
\end{equation}
The viscous stress tensor is assumed to be expressed through the
ordinary relation
\begin{equation} 
\sigma_{i,j} = 
\eta(\phi)\, \dot\gamma_{i,j} ,
\label{sigma}
\end{equation}
with the shear rate tensor
\begin{equation}
\dot\gamma_{i,j}\equiv
{\partial v_i\over\partial x_j}+{\partial v_j\over\partial x_i}
-{2\over 3}\,\delta_{i,j}\,{\partial v_l\over\partial x_l}
.
\label{dot_gamma}
\end{equation}
Note that Eq.(\ref{sigma}) does not represent a linear viscosity because
the viscosity $\eta$ is not constant but depends on the internal state
variable $\phi$ of the medium.

\paragraph{Internal state of the medium:}

\begin{figure}
\centerline{\includegraphics[width=8cm]{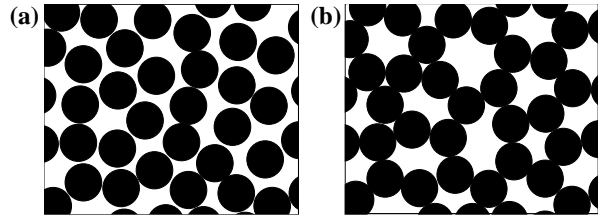}}
\caption{Schematic pictures for granular configurations:
a relaxed state(a) and a jammed state(b).
 }
\label{Gran-conf}
\end{figure}

The dilatant fluid contains dispersed granular particles, which provides
the system with an internal degree of freedom for a macroscopic
description.  Fig.\ref{Gran-conf} shows a schematic illustration for a
relaxed state(a) and that for a jammed state(b).  The internal
state may have a vector or even higher order symmetry in general, but in
this work we study a simple case where the state is represented by a
scalar field $\phi(\mib r)$.  We assign $\phi=0$ for the relaxed state
and $\phi=1$ for the jammed state.

For a given flow field $\mib v(\mib r)$, we assume that there exists a
stationary value $\phi_*$, toward which the state variable $\phi$
changes as
\begin{equation}
\tau\, {D\phi\over Dt} = -\Bigl(\phi - \phi_*\Bigr)
\label{eq.phi}
\end{equation}
with the time scale $\tau$.

We may assume that $\tau$ is constant in the case where the internal
state changes due to the thermal fluctuation or some other mechanism
independent of the flowing field.  
However, we adopt the variable time scale $\tau$ that is inversely
proportional to the local shear rate $\dot\Gamma$,
\begin{equation}
\tau =  r\,\dot\Gamma^{-1}
\label{tau(Gamma)}
\end{equation}
with a dimensionless constant $r$, because it is more natural to suppose
that the state change is driven by the flow deformation.  Note that this
form of $\tau$ does not introduce a new time scale to the system and
makes it respond quite peculiarly to an external impact.  

The stationary value $\phi_*$ is determined by the local flow and we
assume that it is an increasing function of the local stress $S$.
We employ a simple form
\begin{equation}
%
\phi_*(S) = 
\phi_M \cdot {(S/S_0)^2\over 1+(S/S_0)^2 }
\label{phi*(S)}
\end{equation}
with the characteristic shear stress $S_0$.  The parameter $\phi_M$
represents the value of the state variable in the high stress limit and
should depend upon the volume fraction of the granules and some other
parameters of the medium.

For the scalar values of 
the shear rate $\dot\Gamma$ and the shear stress $S$
in Eqs.(\ref{tau(Gamma)}) and (\ref{phi*(S)}),
we adopt the definitions
\begin{equation}
\dot\Gamma \equiv
  \sqrt{{1\over 2}\,\mathrm{Tr}[\hat{\dot\gamma}\cdot\hat{\dot\gamma}]},
\quad
S\equiv \sqrt{{1\over 2}\,\mathrm{Tr}[\hat\sigma\cdot\hat\sigma]},
\end{equation}
which reduce to the ordinary shear rate and shear stress in the case of
simple shear flow.

\paragraph{Viscosity:}

The shear thickening property of the model comes from the
$\phi$-dependence of the viscosity, for which we assume
\begin{equation}
\eta(\phi) = \eta_0 \exp\left[A{\phi\over 1-\phi}\right]
\label{eta_phi}
\end{equation} 
with the viscosity in the relaxed state $\eta_0$ and a dimensionless
parameter $A$.  We have introduced the Vogel-Fulcher type strong
divergence at the jamming point $\phi=1$ in order to represent severe
thickening observed in the dilatant fluid. In Eq.(\ref{eta_phi}), the
state variable $\phi$ plays an analogous role with the inverse
temperature in the glass transition.
Note that the state variable $\phi$ cannot be $\phi>1$ even when
$\phi_M>1$ in Eq.(\ref{phi*(S)}) if we employ Eq.(\ref{tau(Gamma)})
because 
the shear rate vanishes $\dot\Gamma \searrow 0$
as $\phi\nearrow 1$ due to the diverging  viscosity.

\paragraph{Unit system:}

For numerical presentation,
we employ the unit system where
\begin{equation}
\eta_0 = S_0 = \rho = 1,
\label{unit}
\end{equation}
namely, the time, length, and mass are measured by the units
\begin{equation}
\tau_0\equiv {\eta_0\over S_0}, \qquad
\ell_0\equiv \sqrt{{\eta_0\over\rho}\,\tau_0}, \qquad
m_0 \equiv \rho\,\ell_0^3,
\end{equation}
respectively.
The rate $1/\tau_0$ gives the scale for the shear rate where thickening
occurs, and the length scale $\ell_0$ is the corresponding hydrodynamic
length scale.
%
For the cornstarch suspension of 41 wt\%\cite{Bonn-2008}, these
parameters may be estimated as $S_0\approx 50\,{\rm Pa}$, $\eta_0\approx
10\,{\rm Pa\cdot s}$, and $\rho\approx 10^3\,{\rm kg/m^3}$, which give
$\tau_0\approx 0.2\,{\rm s}$ and $\ell_0\approx 5\,{\rm cm}$.



\section{Simple shear flow under external shear stress}
\begin{figure}
\centerline{\includegraphics[width=0.5\textwidth]{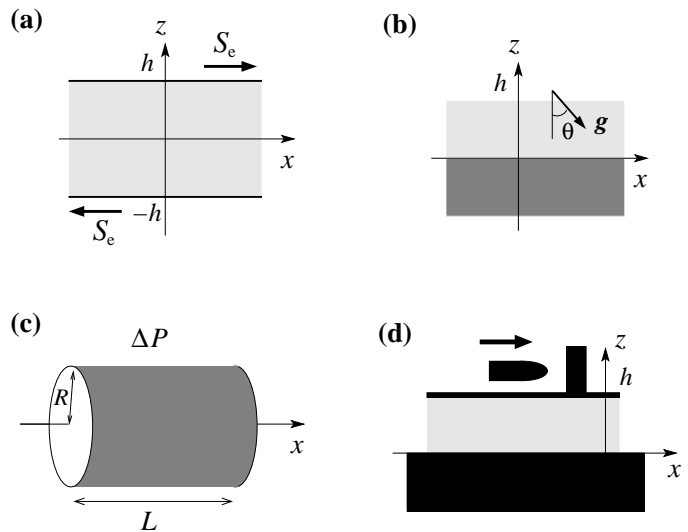}}
\caption{Simple flow configurations and the coordinate system:
(a) shear flow, (b) gravitational slope flow, (c) Poiseuille flow, and
(d) impact by a bullet.}
\label{Flow-conf}
\end{figure}

First, we will study behaviors of the dilatant fluid for a simple shear
flow under an externally applied shear stress(Fig. \ref{Flow-conf}(a)).
The velocity field is assumed to be $\mib v=(u(z,t), 0, 0)$ and the
external stress imposes the boundary condition
\begin{equation}
 S(z,t)\Bigr|_{z=\pm h} = S_{\rm e},
\label{eq-shear-BC}
\end{equation}
where we have introduced the notation for the shear stress
\begin{equation}
S(z,t) \equiv \eta(\phi)\; \dot\gamma(z,t)
\label{S(z,t)}
\end{equation}
with the shear rate
\begin{equation}
\dot\gamma(z,t)\equiv {\partial u(z,t)\over\partial z}.
\label{gamma(z,t)}
\end{equation}
$h$ is the half width of the flow and $S_{\rm e}$ is the applied stress
at the boundaries(Fig. \ref{Flow-conf}(a)).  Then, Eqs.(\ref{eq.v}) and
(\ref{eq.phi}) become
\begin{eqnarray}
\rho{\partial u(z,t)\over\partial t} & = &
 {\partial\over\partial z}S(z,t),
\label{eq-shear-1}
\\
r {\partial\phi(z,t)\over\partial t} & = &
-|\dot\gamma(z,t)|\Bigl(\phi(z,t) - \phi_*\bigl(S(z,t)\bigr)\Bigr),
\label{eq-shear-2}
\end{eqnarray}

In the following, we first examine a steady flow solution, then perform
the stability analysis for the solution and the numerical simulation for
these equations of motion.

\subsubsection{Steady flow solution}

The steady solution for Eqs.(\ref{eq-shear-BC}) $\sim$
(\ref{eq-shear-2}) can be readily obtained as
\begin{equation}
\phi = \phi_*(S_{\rm e}),\quad
\dot\gamma 
  = {S_{\rm e}\over\eta\bigl(\phi_*(S_{\rm e})\bigr)}
  \equiv \dot\gamma_*(S_{\rm e}) .
\label{shear_flow-steady}
\end{equation}
%
\begin{figure}
\centerline{\includegraphics[width=8cm]{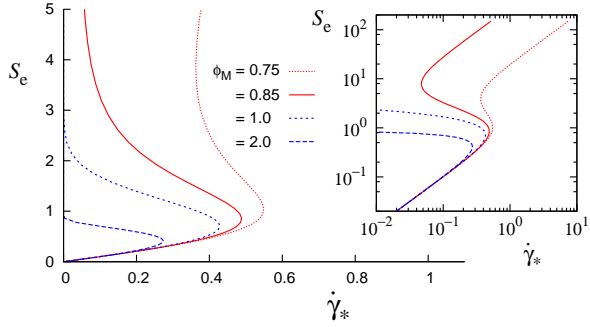}}
%
\caption{(Color online)
The stress-shear rate relation for the viscosity given by
Eq.(\ref{shear_flow-steady}) for various $\phi_M$ with $A=1$.
The inset shows the plots in the logarithmic scale.
}
\label{SF-gamma-S}
\end{figure} 
%
From these equations, we can obtain the relationship between the stress
and the shear rate, which is plotted in Fig.\ref{SF-gamma-S} for various
values of $\phi_M$ with $A=1$.
In the logarithmic plots, the straight line with the slope 1 correspond
to the linear stress-shear rate relation with a constant differential
viscosity.  One can see that there are two regimes: the low viscosity
regime in the low shear stress and the high viscosity regime in the high
shear stress.
Between the two regimes, there is a branch where the shear rate
decreases for increasing shear stress.  The state in the middle branch
can be unstable against infinitesimal perturbation.

From this stress-shear rate relation, we expect there should be
hysteresis upon changing the shear rate; if the system starts from the
low shear rate on the lower branch, the stress increases continuously,
but before the system reaches the end of the lower branch, it should
jump to the upper branch by discontinuous increase of the stress. If the
system starts from the high shear rate on the upper branch and the shear
rate decreases, the stress should jump to the lower branch before the
system reaches the end of the upper branch.
This sudden increase/decrease of stress corresponds to the discontinuous
change of viscosity in the shear thickening.

\subsubsection{Linear stability of the steady flow}

Now, we examine the linear stability of the steady shear flow given by
Eq.(\ref{shear_flow-steady}).  For a full analysis, an arbitrary
perturbation should be allowed, but here we examine the linear
stability against the restricted perturbation where the velocity is in
the $x$-direction and the spatial dependence is only on the $z$
coordinate:
\begin{equation}
\mib v(\mib r,t) = (\dot\gamma_*z + \delta u(z,t), 0, 0),\qquad
\phi(\mib r,t) = \phi_*(S_{\rm e}) + \delta\phi(z,t),
\end{equation}
then the dynamics is analyzed using Eqs.(\ref{eq-shear-1}) and
(\ref{eq-shear-2}).
Even within this restriction, we will see the steady shear flow in the
middle branch may become unstable and the oscillatory flow arises.

The linearized equations for the perturbation are now given by
\begin{eqnarray}
\rho\,{\partial\over\partial t}\delta\dot\gamma(z,t) & = &
 \eta_* {\partial^2\over\partial z^2}\delta\dot\gamma(z,t) + 
\eta_*'\dot\gamma_* {\partial^2\over\partial z^2}\delta\phi(z,t) ,
\\
r\,{\partial\over\partial t}\delta\phi(z,t) & = &
\dot\gamma_*\Bigl(
       \phi_*'\, \eta_* \delta\dot\gamma(z,t) + 
         \bigl(-1+\phi_*' \eta_*' \dot\gamma_*\bigr) \delta\phi(z,t)
            \Bigr) ,
\end{eqnarray}
where the primes denote the derivative by its argument, and we have
introduced the abbreviated notations,
\begin{equation}
\eta_* \equiv \eta\bigl(\phi_*(S_{\rm e})\bigr), 
\quad  
\eta_*' \equiv 
     \left.{d\eta(\phi)\over d\phi}\right|_{\phi=\phi_*(S_{\rm e})}, 
\quad
\phi_*' \equiv {d\phi_*(S_{\rm e})\over d S_{\rm e}}.
\end{equation}
Then, the growth rate $\Omega_k$ of the perturbation for the Fourier
component with the wave number $k$ in the $z$ direction is determined by
\begin{equation}\left|\begin{array}{cc}
\rho\Omega_k + k^2 \eta_*, & k^2 \eta_*' \dot\gamma_* \\
-\dot\gamma_*\phi_*'\, \eta_* , &
 r\Omega_k - \dot\gamma_*\bigl(-1+\phi_*' \eta_*' \dot\gamma_*\bigr)
\end{array}\right| = 0.
\end{equation}

This gives a positive real part of $\Omega_k$ for the wave number $k$
that satisfies
\begin{equation}
0<  k^2 <  k_c^2
\equiv 
{1\over r}\,\left({\rho\over\eta_*}\right)
         S_{\rm e} \left(-{d\dot\gamma_*\over d S_{\rm e}}\right)
\label{shear-flow-stability}
\end{equation}
in the case $d\dot\gamma_*/d S_{\rm e}<0$, i.e. $S_{\rm
e}$ is in the unstable branch of the shear stress-shear rate curve.
%
%
Since the smallest possible wave number $k$ for the perturbation is
$\pi/(2h)$ and $\eta_*/\rho$ is the kinematic viscosity for the steady
flow, we can interpret this result in the way that {\em the steady shear
flow in the unstable branch is unstable as long as the system width is
larger than the momentum diffusion length due to the viscosity.}

For a given external shear stress $S_{\rm e}$ in the unstable branch, the flow
becomes unstable for the system wider than $2h_c\equiv \pi/k_c$, where
the growth rate $\Omega_k$ has a finite imaginary part $\omega_c$ given
by
\begin{equation}
\omega_c \equiv \sqrt{S_{\rm e}\over r\rho}\; k_c =
 {1\over r}
   \sqrt{S_{\rm e}\left(-{d\dot\gamma_*\over dS_{\rm e}}\right) \dot\gamma_*}.
\end{equation}
%
Note that the scales of $k_c$ and $\omega_c$ are typically set by
$1/\ell_0$ and $1/\tau_0$ although their actual values depend on $r$ and
the other system parameters, i.e. $A$ and $\phi_M$.

\subsubsection{Shear thickening oscillation in unstable shear flow}

\begin{figure}
\centerline{\includegraphics[width=7cm]{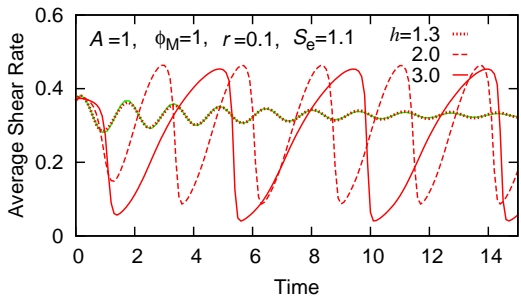}}
\caption{(Color online)
Oscillation of the average shear rate $u(h)/h$ in the shear
flow for $h=1.3$, 2, and 3 with $A=\phi_M=1$, $r=0.1$, and $S_{\rm e}=1.1$.
The (green) line, that overlaps the plot for $h=1.3$, shows the
plot for $f(t)=c_1+c_2\, e^{-t/\tau}\sin(\omega t+\theta)$ with
$\omega=4.0$, $\tau=5.8$, $\theta=0.86$, $c_1=0.33$, and $c_2=0.056$.
}
\label{SF-SurfV}
\end{figure}

The oscillatory behavior of the shear flow in the unstable regime can be
seen by numerically integrating Eqs.(\ref{eq-shear-1}) and
(\ref{eq-shear-2}) with Eqs.(\ref{S(z,t)}) and (\ref{gamma(z,t)}).
In Fig.\ref{SF-SurfV}, the average shear rates $u(h)/h$ for
(anti-)symmetric solutions are plotted as a function of time for various
system width $h$ with the constant shear stress $S_{\rm e}=1.1$ in the
unstable regime for $A=\phi_M=1$ and $r=0.1$. The initial state is
prepared as the steady solution (\ref{shear_flow-steady}) for $S_{\rm e}=1$.
For this set of parameters, $k_c=1.18$, which gives $h_c=1.33$ and
$\omega_c=3.91$.  

For $h=1.3$, which is smaller than $h_c$, the flow shows overdumped
sinusoidal oscillation with the angular frequency 4.0, which is close to
$\omega_c$.  For larger $h$, the oscillation becomes self-sustained and
non-linear; the gradual buildups of the flow speed are followed by
sudden drops.

\begin{figure}
\centerline{\includegraphics[width=8cm]{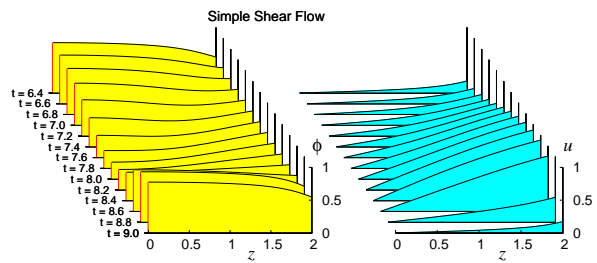}}
\caption{(Color online)
Time development of $\phi(z)$ (left) and $u(z)$ (right) in the
shear flow oscillation for  $A=\phi_M=1$, $r=0.1$, $S_{\rm e}=1.1$, and $h=2$.
Only the positive parts of the flow ($z>0$) are presented.
}
\label{SF-t-develop}
\end{figure}

This non-linear oscillation of shear thickening fluid is shown in more
detail in Fig.\ref{SF-t-develop}, where the time development of the
$\phi(z)$ and $u(z)$ are plotted.  Only the positive half of the
solution is plotted for the (anti-)symmetric solution.  In the plots,
the oscillation starts from the almost uniform shear flow in the high
viscosity state with a larger value of $\phi$.  This flow cannot be
completely uniform because the external shear stress $S_{\rm e}$ is in the
unstable branch, thus the flow speed builds up gradually as the internal
state $\phi$ relaxes to reduce the viscosity, but eventually, $\phi$
starts increasing when the shear stress becomes large enough.  Then,
larger value of $\phi$ causes higher viscosity, which decelerates the
flow speed, but this causes even higher value of $\phi$ because the
inertia stress due to the deceleration is added on the top of the stress
by the shear flow, which results in the sudden drop of the flow speed.


\section{Gravitational slope flow and Poiseuille flow}

Similar analyses are performed for a gravitational slope flow and
Poiseuille pipe flow.

\subsection{Gravitational slope flow}

For the gravitational slope flow (Fig.\ref{Flow-conf}(b)),
Eqs.(\ref{eq.v}) and (\ref{eq.phi}) should be solved with the boundary
conditions
\begin{equation}
\mib v\bigr|_{z=0}=0,\qquad
\hat\sigma\cdot\mib n\bigr|_{z=h(x,y)} = 0,
\label{BC-GF}
\end{equation}
where we have assumed that the bottom of the flow is located at $z=0$
and the flow depth at $(x,y)$ is given by $h(x,y)$; the vector $\mib n$
represents the normal vector to the flow surface.  The gravitational
body force is given by
\begin{equation}
\mib g = ( g\sin\theta,0, -g\cos\theta)\equiv (g_\parallel,0, -g_\perp)
\end{equation}
with  the slope angle $\theta$.


For the flow field $\mib v=(u(z,t),0,0)$, Eqs.(\ref{eq.v}) and
(\ref{eq.phi}) become
\begin{eqnarray}
\rho\;{\partial u(z,t)\over\partial t} & = & 
{\partial  S(z,t)\over\partial z}
  + g_\parallel ,
\label{eq.grav-1}
\\
0 & = &  -{\partial P(z)\over\partial z} -  g_\perp ,
\label{eq.grav-2}
\\
r{\partial\phi(z,t)\over\partial t} & = & 
-|\dot\gamma(z,t)|\Bigl(\phi(z,t) - \phi_*\bigl(S(z,t)\bigr) \Bigr)
\label{eq.grav-3}
\end{eqnarray}
with the shear stress (\ref{S(z,t)})
and the boundary conditions (\ref{BC-GF}) are given by
\begin{equation}
u(0)=0, \qquad \left.{\partial u(z,t)\over\partial z}\right|_{z=h} = 0.
\label{eq.grav-bc}
\end{equation}
Eq.(\ref{eq.grav-2}) can be solved immediately to give the pressure
\begin{equation}
 P(z) = g_\perp (h-z)+ P_0
\end{equation}
with the atmospheric pressure $P_0$.

The steady solution for Eqs.(\ref{eq.grav-1}) and (\ref{eq.grav-3}) 
under the boundary condition (\ref{eq.grav-bc}) is given by
\begin{equation}
\phi(z) = \phi_*\bigl(S_g(z)\bigr), \quad
\dot\gamma(z) = {S_g(z)\over\eta\Big(\phi_*\bigl(S_g(z)\bigr)\Big)}
\end{equation}
with the gravitational shear stress
\begin{equation}
S_g(z) \equiv  g_\parallel (h-z).
\end{equation}
From these,
the flow speed $u(z)$ and the flux per unit width $\Phi_G$ can be
calculated by
\begin{equation}
u(z)  =  \int_0^z \dot\gamma(z')\, dz' , \quad
\Phi_G  =  \int_0^h u(z) dz .
\label{gravi-u-Phi_G}
\end{equation}

\begin{figure}
\centerline{
\includegraphics[width=8cm]{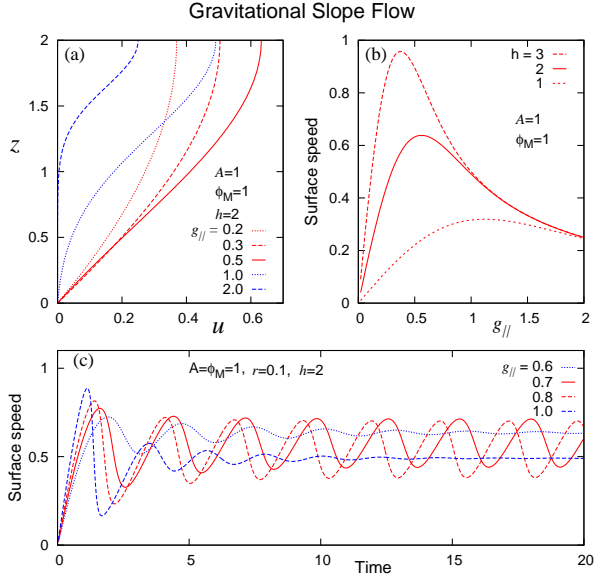}
} 
\caption{(Color online)
 Steady Gravitational flows:
(a) the flow speed profiles as a function of $z$, 
(b) the surface flow speed vs $g_\parallel$, and
(c) the time development of the surface speed.
}
\label{GF-u}
\end{figure}

In Fig.\ref{GF-u}, the flow speed profiles and the surface speeds given
by Eq.(\ref{gravi-u-Phi_G})
plotted for several sets of parameters.
The depth dependences of the flow speed are shown in Fig.\ref{GF-u}(a)
for some values of $g_\parallel$; For small $g_\parallel$, the flow
speed depends upon the depth parabolically as in a Newtonian fluid,
while, for larger $g_\parallel$, the flow speed profile develops a
convex part, which corresponds with the unstable branch of
Fig.\ref{SF-gamma-S} in the shear flow.  In Fig.\ref{GF-u}(b), the
surface speed $u(h)$ are plotted as a function of $g_\parallel$ for some
values of $h$. One can see they decreases for large $g_\parallel$, which
means that the fluid flows slower for larger inclination angle.  This is
because the viscosity of the fluid becomes large in the high shear
stress caused by large $g_\parallel$.

\subsection{Poiseuille Flow}

Pressure driven pipe flow with the cylindrical symmetry around the
$x$-axis (Fig.\ref{Flow-conf}(c)) is governed by the equation
\begin{eqnarray}
\rho\,{\partial u(r,t)\over\partial t} & = &
+{\Delta P\over L} + 
{1\over r}\,{\partial\over\partial r}\Bigl(r S(r,t)\Bigr),
\label{eq.pois-1}
\\
r{\partial\phi(r,t)\over\partial t} & = & 
-|\dot\gamma(r,t)|\Bigl(\phi(r,t) - \phi_*\bigl(S(r,t)\bigr) \Bigr)
\label{eq.pois-3}
\end{eqnarray}
with the shear stress and the shear rate
\begin{equation}
S(r,t) = \eta(\phi)\dot\gamma(r,t),\quad
\dot\gamma(r,t)= {\partial u(r,t)\over\partial r}.
\end{equation}
Here, $\Delta P$ ($>0$) is the pressure drop along the pipe over the
length $L$, and $r$ is the distance from the central axis:
$r\equiv\sqrt{y^2+z^2}$.

The steady flow solution for this configuration is given by
\begin{equation}
\phi(r) = \phi_*(S_P(r)),\quad
\dot\gamma(r) = 
    {S_P(r)\over\eta\Bigl(\phi_*\bigl(S_P(r)\bigr)\Bigr)}
\end{equation}
with the Poiseuille shear stress
\begin{equation}
S_P(r) \equiv -\,{1\over 2}\,{\Delta P\over L} r.
\end{equation}

\begin{figure}
\centerline{\includegraphics[width=8cm]{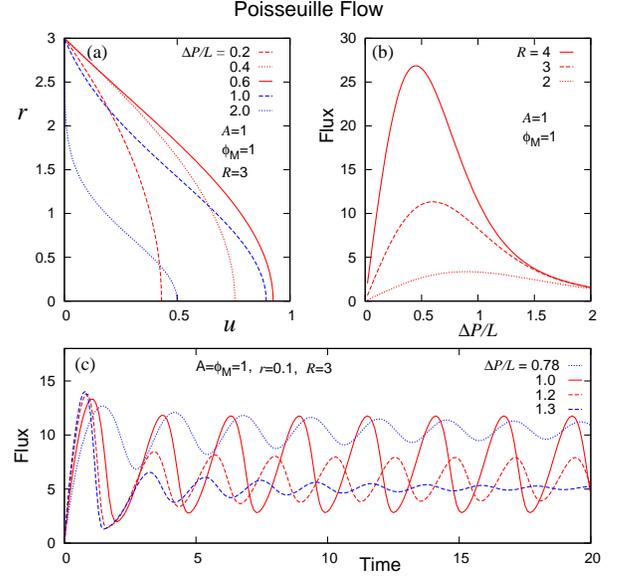}
}
\caption{(Color online)
Poiseuille flow: 
(a) the flow speed profiles as a function of $r$, 
(b) the flow flux vs pressure gradient $\Delta P/L$, and
(c) the time development of the flux.
}
\label{PF-flux}
\end{figure}

In Fig.\ref{PF-flux}, the flow speed profiles $u(r)$ and the flow flux
$\Phi$ defined as
\begin{equation}
\Phi \equiv \int_0^R u(r) 2\pi r\, dr
\end{equation}
are plotted.  
General features of the flow is analogous to those of the gravitational
flow, and
the flow flux decreases upon increasing the pressure
gradient for the large pressure gradient because of the shear thickening.

\subsection{Shear thickening oscillation in gravitational flow and 
Poiseuille flow}

These steady flows become unstable when the shear stress is in the range
of the unstable branch at some region of the flow.  The oscillations in
the surface flow speed and the flow flux are plotted for the
gravitational and Poiseuille flow in Figs.\ref{GF-u}(c) and
\ref{PF-flux}(c), respectively.  The shear thickening oscillation
appears in a large enough system for a certain range of external drive
$g_\parallel$ or $\Delta P/L$; The system length scale should be larger
than the viscous length scale of the flow, and the external drive should
be in the range where some part of the flow is in the unstable branch.
From the plots, one can see the oscillation disappears when the external
drive is either too small or too large.  
In the former case, the fluid behaves as Newtonian while in the latter
case the size of the unstable region becomes too small.
The shape of the oscillation in
the non-linear oscillation regime is saw-teeth like, i.e. gradual
increases followed by sudden drops, as we have discussed in the simple
shear flow case.

\begin{figure*}
\centerline{
\includegraphics[width=8cm]{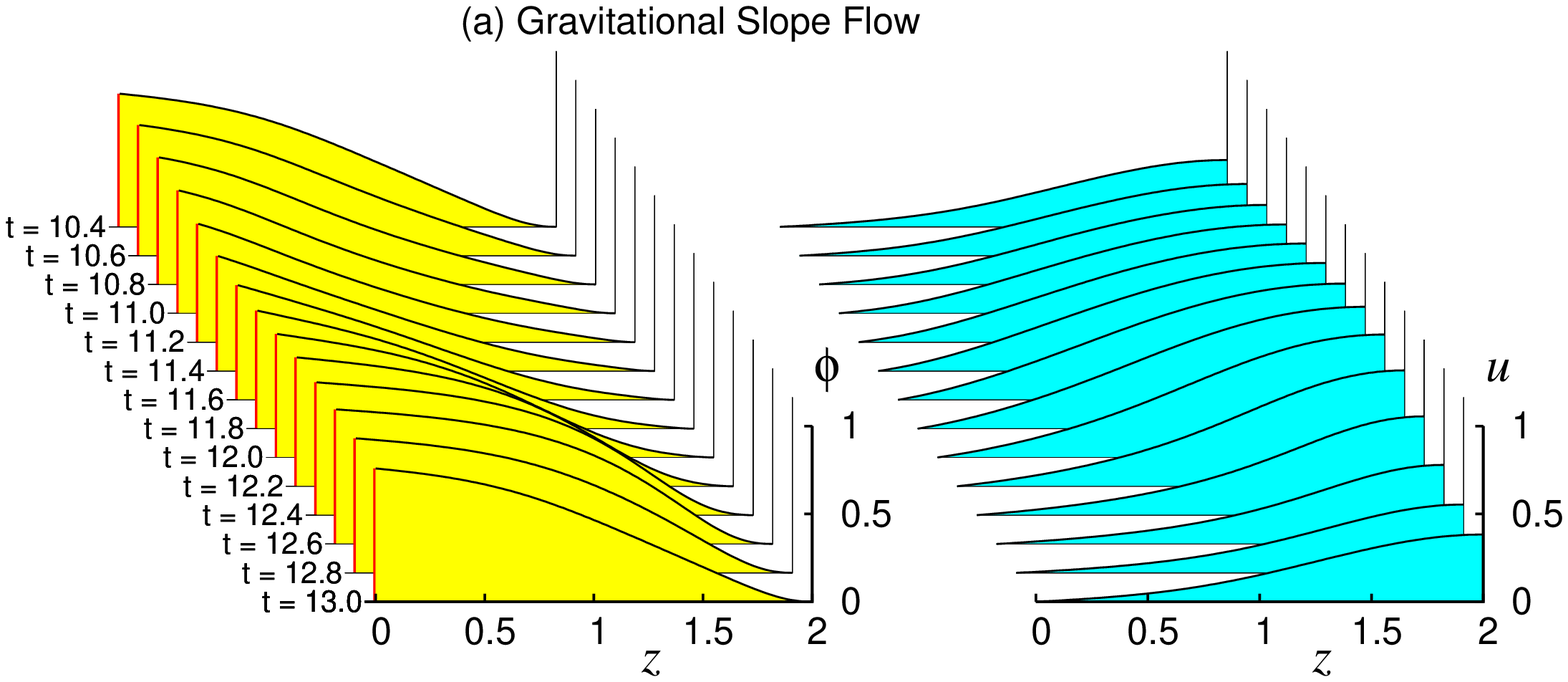}
\includegraphics[width=8cm]{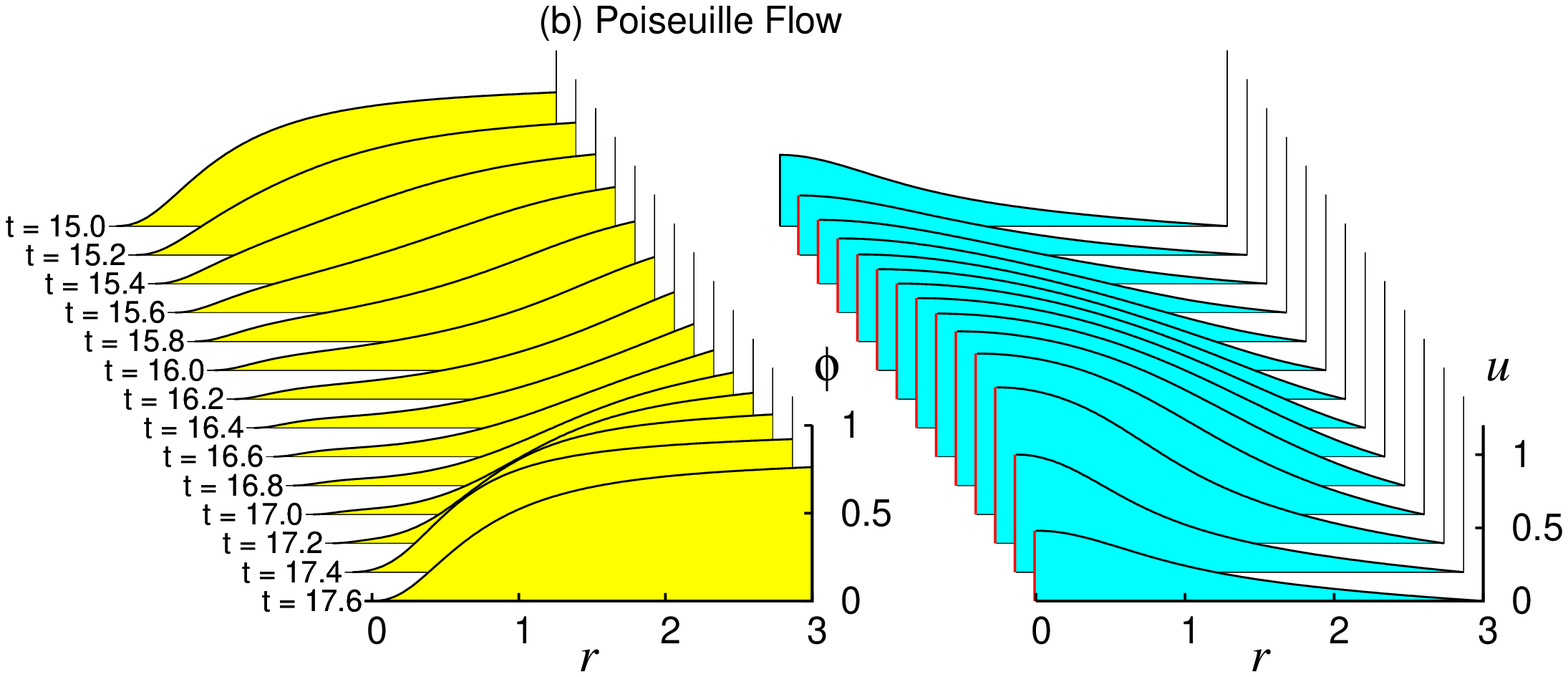}
} 
\caption{(Color online)
 Time development of $\phi(z)$ (left) and $u(z)$ (right) in the
oscillatory flow of the
gravitational slope flow (a) and the Poiseuille flow (b).
The parameters are
$A=\phi_M=1$ and $r=0.1$ with $g_\parallel=0.8$ and $h=2$ for the
 gravitational flow and with $\Delta P/L=1$ and $R=3$ for Poiseuille flow.  }
\label{GF-PF-t-u-phi}
\end{figure*}

The spatial variation of oscillatory flow are shown in
Fig.\ref{GF-PF-t-u-phi}.  The general feature is the same with that of
the shear flow, but the value of $\phi$ is zero at the surface of the
gravitational flow and at the center of Poiseuille flow because the
shear is zero.

\section{Response to an external impact}

One of the peculiar features of the dilatant fluid is instantaneous
hardening by an external impact. It hardens almost immediately upon
application of an external impact and allows little deformation like
rigid material.  It has been demonstrated that the hardening is so rapid
that the material can be used for a body armor to stop a
bullet\cite{Wagner-2009}.  Such instantaneous hardening cannot be
explained by the transformation between steady configurations of
granules, but must be a result of the failure to rearrange the granular
configuration due to some obstruction.
Upon sudden impact, the granules are inhibited to rearrange their
configurations due to to either dissipation by the interstitial fluid or
the jamming by direct contacts.
In the case of slow deformation, the stress is low and the lubrication
due to the fluid allows the granules to re-arrange themselves so that
they can pass each other.

In the present model, this aspect of the medium is represented by 
Eq.(\ref{tau(Gamma)}) that the relaxation rate of the internal state is
proportional to the shear rate.  For a sudden deformation, the state
variable changes to a high stress value as the medium deforms; When the
medium is dense ($\phi_M\gtrsim 1$) and the external impact is strong
enough, the state reaches the jammed state after a certain amount of
deformation, which is almost independent of the speed of deformation.

In order to demonstrate this aspect of the model, we perform simple
simulations that the layer of fluid of the thickness $h$ is driven by
a sudden motion of the upper boundary wall at $z=h$ with the fixed lower
boundary at $z=0$ (Fig.\ref{Gran-conf}d).
Let $U(t)\equiv u(h,t)$ be the velocity of the upper wall.
Initially, the fluid is at rest,
\begin{equation}
u(z,t) = 0, \quad \phi(z,t)=0, \quad U(t)=0
\quad\mbox{for $t<0$},
\end{equation}
then the upper wall is moved suddenly by the velocity $u_0$ at $t=0$,
$U(0)=u_0$.  For $t>0$, the velocity of the upper wall is determined by
\begin{equation}
m{dU(t)\over dt} =
 -\eta\bigl(\phi(h,t)\bigr){\partial u(z,t)\over\partial z}\Bigr|_{z=h},
\end{equation}
with Eqs.(\ref{eq-shear-1}) and (\ref{eq-shear-2}), where $m$ is the
mass of the upper wall per unit length.

\begin{figure}
\centerline{\includegraphics[width=8cm]{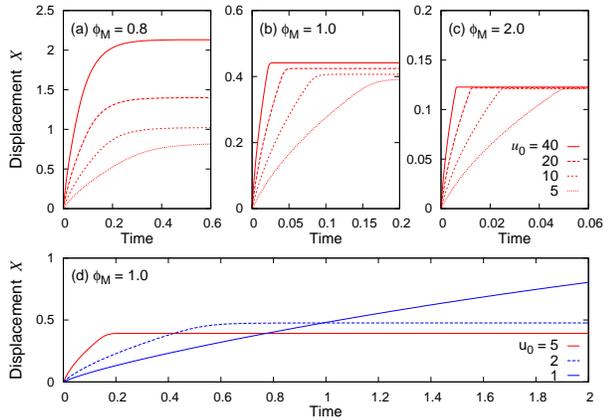}
} \caption{(Color online)
The time dependence of the displacement $X$ after the impact for the
system of $\phi_M=0.8$ (a), 1 (b), and 2 (c) with the initial speed
$u_0=40$, 20, 10, and 5, and for the system of $\phi_M=1.0$ with
$u_0=5$, 2, and 1 (d).  The other parameters are $h=2$, $r=0.1$, and
$A=m=1$.
}
\label{impact}
\end{figure}

Fig.\ref{impact} shows the displacement $X(t)$ of the upper wall,
\begin{equation}
X(t) = \int_0^t U(t') dt' ,
\end{equation}
for the three cases, $\phi_M=0.8$, 1, and 2 for various initial speeds
$u_0$ increases.  The wall decelerates rapidly as the fluid thickens in
response to the stress, and eventually stops. For $\phi_M=0.8$, the
final wall displacement increases as the initial speed $u_0$.  On the
other hand, for $\phi_M=1$ and 2, the final displacement hardly depend
on $u_0$ when $u_0>5$.  This is because the fluid gets jammed
at a certain strain as it deforms, and cannot deforms further.  However,
when the initial speed is small enough, the upper wall does not stop
quickly because the fluid does not thicken as shown Fig.\ref{impact}(c).

\section{Two dimensional inhomogeneous flow}

Now,
we present the results of numerical simulations for two
dimensional system in the simple shear configuration
(Fig.\ref{Flow-conf}(a)) in order to examine how the inhomogeneity in
the $x$  direction affects the system behavior, especially in the
case of shear thickening oscillation.

The velocity field is assumed to be in the $x-z$ plane, $\mib
v=(u(x,z,t), 0,w(x,z,t))$, and in the $x$ direction we employ the
periodic boundary condition with the system length $L$.
We take $L=10h$ in the present simulations.

The fluid dynamic equation (\ref{eq.v}) is integrated using the standard
MAC(Marker-and-Cell) method\cite{MAC} for the incompressible fluid, and
$|\mib\nabla\cdot\mib v|$ is kept less than $10^{-10}$.  
Euler method is employed for
the time integration of Eqs.(\ref{eq.v}) and (\ref{eq.phi}).

The motion of the plates at $z=\pm h$ is controlled so that the average
shear stress on the plate is equal to $S_{\rm e}$,
\begin{equation}
{1\over L}
\int_0^L \eta\bigl(\phi(\mib r,t)\bigr)\dot\gamma_{xz}(\mib r, t)
\Bigr|_{z=\pm h} dx  = S_{\rm e}.
\end{equation}

As for the initial configuration at $t=0$, we assume that the fluid is
at rest and the state variable $\phi$ is close to
zero with small fluctuations introduced at every computational grid
point $\mib r_i$,
\begin{equation}
\mib v(\mib r,0)=0, \quad
\phi(\mib r_i, 0) = \xi_i,
\end{equation}
where $\xi_i$ is a random variable uniformly distributed over
$[0,\epsilon)$ with  a small parameter $\epsilon$.  We take
$\epsilon=10^{-4}$.  Note that, for the case of $\epsilon=0$, all
quantities do not depend on $x$, thus the simulations reduce to the one
dimensional case given in Sec. III.

\subsection{Flow diagrams}

\begin{figure}
\centerline{\includegraphics[width=8cm]{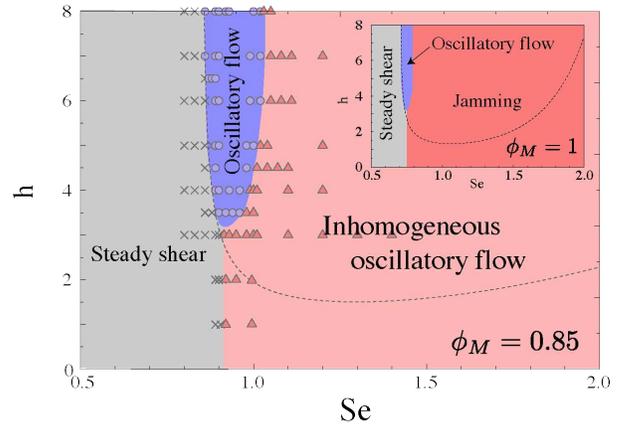}} 
\caption{(Color online)
Flow diagram of the shear flow for $\phi_M=0.85$. The
internal state variable $\phi$ does not depend on $x$ in the region
colored with gray and blue. The inset is the same diagram obtained for
$\phi_M=1$. The other parameters are $A=1$, $r=0.1$, with $L=10h$.
}

\label{Flow-diag}

\end{figure}

Fig.\ref{Flow-diag} shows a flow diagram in the $S_{\rm e}-h$ plane for
$\phi_M=0.85$ with $A=1$ and $r=0.1$ (the inset for $\phi_M=1$).  The
diagrams are determined by the simulations at the points with
marks.

\begin{figure}
\centerline{\includegraphics[width=8cm]{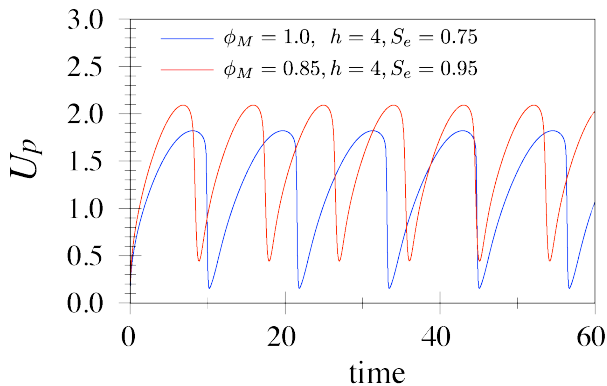}}
\caption{(Color online)
 Time development of the upper plate velocity $U_p$ in the
oscillatory flow regime in the two-dimensional simulation.  The initial
fluctuations decay quickly and the flows show homogeneous oscillation as
 in the case of the one-dimensional system.
}

\label{fig_osc}

\end{figure}

In the steady shear flow (grey) and the oscillatory flow (purple)
regions, the initial fluctuations do not grow, thus the flows remain
homogeneous in the $x$ direction and are the same with those in the
corresponding one dimensional cases in Sec.III (Fig.\ref{fig_osc}); the
dashed lines show the boundary for the two regimes in the
one-dimensional case given by
\begin{equation}
k_c(S_{\rm e}) = {\pi\over 2h}
\end{equation}
using the definition of $k_c$ in Eq.(\ref{shear-flow-stability}) as a
function of $S_{\rm e}$.  In the low $S_{\rm e}$ side, one can see that this
coincides with the corresponding boundary in the two-dimensional case
between the steady shear flow and the oscillatory flow.

The major difference between the one and the two dimensional cases is
that these two homogeneous flow regimes are limited to the smaller $S_{\rm e}$
side. In the larger $S_{\rm e}$ region, the initial fluctuations in the state
variable $\phi$ grows, thus the flow results in the inhomogeneous flow
in the case of $\phi_M=0.85$(pink) or the jammed flow in the
case of $\phi_M=1.0$ of the inset.
In the following, we examine the flows in these two regimes.


\subsection{Inhomogeneous oscillatory flow}

\begin{figure}
\centerline{\includegraphics[width=8cm]{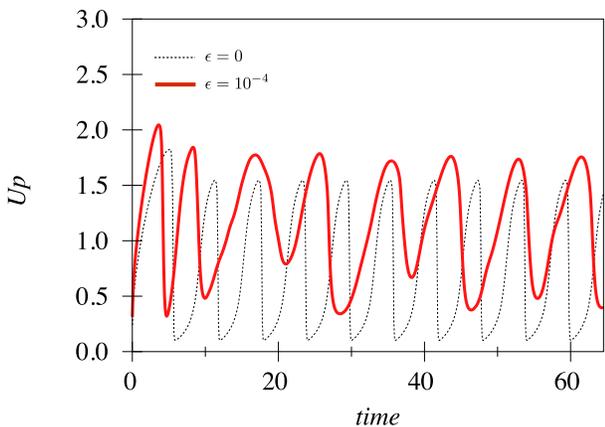}} 
\caption{(Color online)
 The time evolution of the upper plate velocity $U_p$ in the
inhomogeneous oscillatory flow regime with $\phi_M=0.85$.  The other
parameters are $h=5$, $L=50$, $S_{\rm e}=1.5$, $A=1$, and $r=0.1$.
The initial fluctuation is given by $\epsilon=10^{-4}$.  The uniform
oscillation flow with $\epsilon=0$ (the dashed line) is shown for comparison.
}
\label{inhomo-U_p-osci}

\end{figure}

First, we examine the flow for $\phi_M=0.85$.  In this case, the
viscosity does not diverge and the medium keeps flowing.
In Fig.\ref{inhomo-U_p-osci}, the time evolution of the upper plate
velocity $U_p$ is plotted
%
%
along with the case without initial fluctuations.  The flow shows
irregular oscillation with smaller amplitude compared with the noiseless
case.

\begin{figure}
\centerline{\includegraphics[width=8cm]{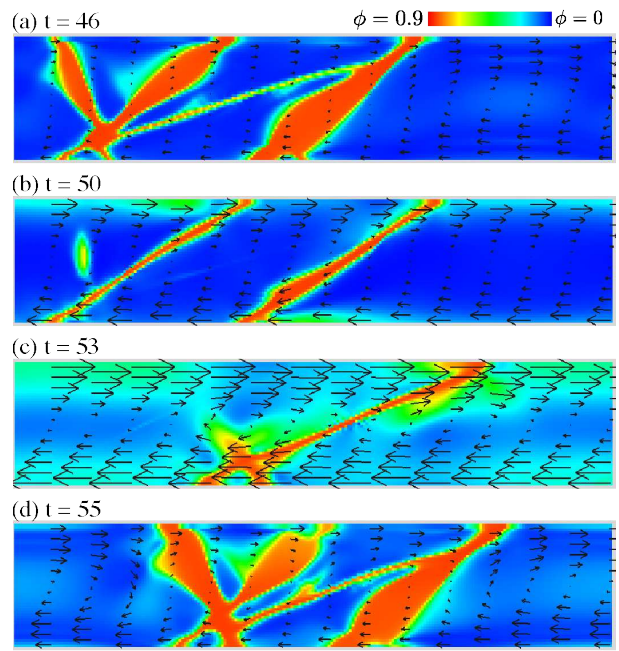}} 
\caption{(Color online)
 The snapshots of the state variable $\phi$ taken during a
cycle of oscillation presented in Fig.\ref{inhomo-U_p-osci}. The arrows
indicate flow velocity. } \label{inhomo-osci-phi}
\end{figure}

The snapshots of $\phi$ for a single cycle of oscillation in
Fig.\ref{inhomo-osci-phi} reveal that the whole system is not thickened
and the oscillation is governed by a few thickening bands.
At the time when $U_p$ reaches its minimum (a), the thickening branch
with the high value of $\phi$ (the red region) is being extended along
the direction of $(1,1)$. 
As the system flows, this branch is stretched and $U_p$ gradually
increases (b),
and eventually the branch breaks off  and $U_p$ reaches maximum (c).
Then, high shear rate makes the broken branches extend again to the other
side to cause sudden deceleration (d).
The thickening branches first appear both in $(1,1)$ and $(1,-1)$
direction, but the latter tend to disappear and transforms into the
$(1,1)$ direction in the course of time, and only the thickening
branches in the $(1,1)$ direction remain.


\subsection{Jamming caused by the instability}

For the case of $\phi_M=1$, the viscosity can diverges and the
instability of the homogeneous flow causes the jamming to stop the flow.

\begin{figure}
\centerline{\includegraphics[width=7.5cm]{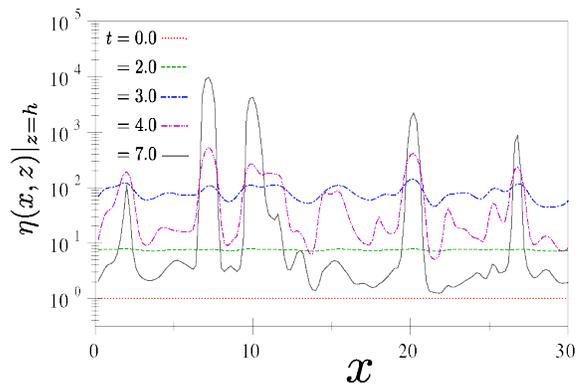}} 
\caption{The spatial variation of viscosity $\eta$ at $z=0$ at several
times in the jamming regime. The parameters are $h=3$, $L=30$, $S_{\rm
e}=1.5$, $\phi_M=1$ with $A=1$ and $r=0.1$.}

\label{inhomo-jam}
\end{figure}
\begin{figure}
\centerline{\includegraphics[width=8cm]{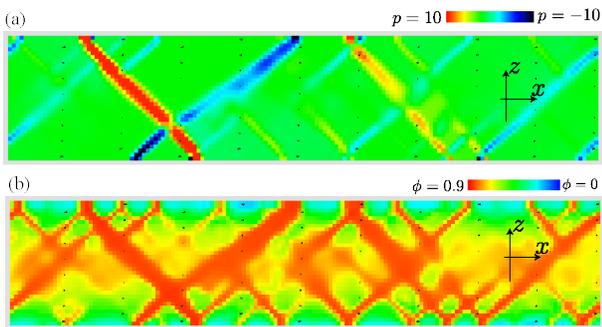}} 
\caption{(Color online)
(a) The spatial distribution of the pressure $P$ and (b) the
internal state variable $\phi$ in the system presented in
Fig.\ref{inhomo-jam} at t = 4. }

\label{inhomo-jam-b}
\end{figure}
\begin{figure}
\centerline{\includegraphics[angle=0,width=9cm]{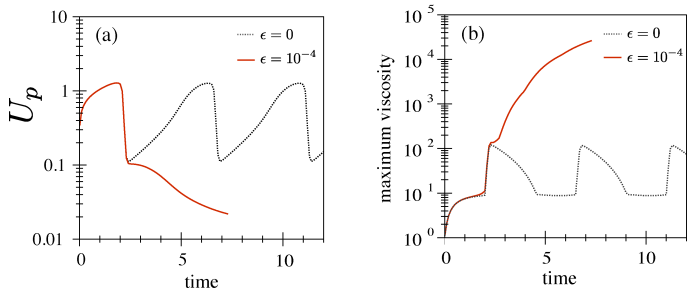}} 
\caption{(Color online)
The time evolution of (a) the upper plate velocity $U_p$, and
(b) the maximum viscosity in the system presented in
Fig.\ref{inhomo-jam}. The dashed lines represent the oscillatory flow
without fluctuations.  }

\label{inhomo-U_p-jammed}
\end{figure}

Figs.\ref{inhomo-jam} and \ref{inhomo-jam-b} shows the simulation
results for $\phi_M=1$.
%
%
The time evolution of the viscosity
distribution at $z=0$ is shown in Fig.\ref{inhomo-jam}.  Initially, the
viscosity is rather uniform with some fluctuations, but peak structure
appears soon around $t=3$ with a certain characteristic length
scale. Some of the peaks grow sharply and the thickening regions
strongly localize($t\gtrsim 4$), then the system is jammed and the flow
stops within a period of oscillation of the homogeneous
case(Fig.\ref{inhomo-U_p-jammed}).


Fig.\ref{inhomo-jam-b} shows the color maps of the pressure $P$ (a) and
the state variable $\phi$ (b) at $t=4$. The fluctuation of $\phi$ first
stands out just below the moving plates, but higher $\phi$ regions form
band structure, and extend along the principal axis of the shear
deformation, $(1,1)$ and $(1,-1)$.  Some of the bands reach the upper
plate from the lower plate, and they jam the system.

In Fig.\ref{inhomo-U_p-jammed}(a), we present the velocity of upper
plate $U_p$ as a function of time. The solid line shows the time
evolution starting from the internal state with fluctuations, and the
dotted line represents the case without fluctuations for comparison.
The state variable suddenly loses homogeneity at $t=2.3$, then
thickening branches appears, and the velocity $U_p$ drops to zero. The
maximum viscosity is always found inside the thickening branch for
$t\gtrsim 2.3$, and its value sharply increases as plotted in
Fig.\ref{inhomo-U_p-jammed}(b).
We cannot simulate the system up to the time when the plates motion
actually stops because the numerical time integration becomes difficult
as the viscosity becomes large, since it requires smaller time step.
%
In the present case, however, we expect the system is jammed because the
decrease of $U_p$ and the increase of maximum viscosity are rapid and
monotonic.

\section{Summary and Discussions}

The shear thickening shown by a dense mixture of granules and fluid has
some peculiar features: (i) instantaneous hardening, (ii) fast
relaxation to flowing state, (iii) rigid thickened state, (iv)
hysteretic thickening transition, (v) oscillatory flowing behavior.  We
constructed a fluid dynamics model by introducing a phenomenological
state variable and showed that the model can describe these features.
Especially, we demonstrated that {\em the shear thickening oscillation}
appears in various shear flow configurations.

\paragraph{Comparison with visco-elasticity:}

A visco-elastic fluid such as a polymer melt shows analogous behavior to
the dilatant fluid; It behaves like a solid in a short time scale and
like a fluid in a long time scale.  This may be compared with that of
the dilatant fluid, i.e. instantaneous hardening in response to an
external impact and fluidization after relaxation of the applied stress.
However, there are some important differences; the visco-elastic medium
changes its behavior according to the observation time scale, and it
also allows large elastic deformation even in a short time solid like
behavior.  On the other hand, the dilatant fluid changes its behavior
according to the stress, i.e. it stays hardened while it is under the
stress and starts flowing within a few second after the removal of the
applied stress. The dilatant fluid allow little deformation even
under large stress.

\paragraph{Shear stress thickening:}

In constructing the model, we assume the fluid is {\em shear-stress
thickening}, i.e. the viscosity depends upon the state variable $\phi$,
and the steady value of the state variable $\phi_*$ is determined by the
local stress as in Eq.(\ref{phi*(S)}).
It is instructive to see what would happen if we assume $\phi_*$ as
a function of the shear rate $\phi_*(\dot\gamma)$.  In this case, the
viscosity is directly given as a function of the shear rate,
$\eta\bigl(\phi_*(\dot\gamma)\bigr)$, thus we should not have a
discontinuous thickening unless we assume a discontinuity  either in
$\phi_*(\dot\gamma)$ or in $\eta(\phi)$.
%
%

Experimentally, the most direct evidence for the shear-stress thickening
should be obtained by the observation that the pipe flow flux is not
monotonically increasing as a function of the applied pressure gradient.

\paragraph{State variable:}

Although we introduced the state variable $\phi$ phenomenologically, we
suppose that the variable represents a certain microscopic property of
the medium, such as contact numbers between grains, associated with the
restrictions against local rearrangement of granular configuration.  The
variable could be a vector or a tensor, but we examined the scalar case
for simplicity.  It is notable that even the scalar state variable
produces the anisotropic stress chain like structure in the system as we
have seen in the two-dimensional simulations.

\paragraph{Jamming and response to impact:}

A remarkable feature of the dilatant fluid is that hardening response is
so instantaneous that the medium can be used for a body armor to stop a
bullet\cite{Wagner-2009}.  We believe that such an instantaneous severe
hardening cannot be explained by a transformation between two steady
states, i.e. from the fluid state under low stress to the rigid state
under high stress.  Instead, it must be a result that the rapid
rearrangement in the granular configuration is inhibited.
There are two possible mechanisms that inhibit the granular
rearrangement in the densely packed medium: the Reynolds dilatancy and
the formation of stress chains.  The Reynolds dilatancy can inhibit
rearrangement by the fluid friction because the rearrangement should
induce the strong interstitial fluid flow through granules in order to
compensate the local volume change caused by the dilatancy.  The stress
chains through direct contacts between granules can be formed by the
application external stress and prevent granules from being rearranged.

In the present model, such an aspect of hardening is represented by the
fact that the relaxational time scale for the state change is not
constant but proportional to the shear rate (\ref{tau(Gamma)}).  Then,
the state variable $\phi$ reaches a steady value $\phi_*$ in a time
scale where the strain changes by the amount $r$.  Consequently, the
medium with $\phi_M\gtrsim 1$ can deform only up to a certain strain
around $r$ by a hard impact.

\paragraph{Shear thickening oscillation:}

One of interesting results of the present model is the shear thickening
oscillation.  The steady shear flow is unstable when the flow is in the
unstable branch of the shear stress-shear rate curve and is wide enough
compared to the diffusion length scale by the viscosity. In this case,
the flow shows the oscillation between the thick state in the high shear
regime and the thin state in the low shear regime.  In two dimensions,
uniform oscillation appears only in the smaller external shear stress
region, but some oscillatory behavior remains even when inhomogeneity
develops in the flow.

The shear thickening oscillation in the homogeneous flow looks
similar to the stick-slip motion in a frictional system.  However, 
there are some important differences; the stick-slip motion starts with
the sudden acceleration caused by the slip weakening resistance under
the constant speed driving through a mechanism with finite rigidity,
while the shear thickening oscillation starts with the sudden
deceleration caused by the shear thickening transition under the
constant force driving.

%
Aradian and Cates have also studied the dynamics of shear thickening
fluids and found oscillatory behaviors\cite{Aradian-2006}.  They
focused, however, on the regime where the structural relaxation time is
much larger than the fluid dynamical time scale, which is appropriate
for liquid crystal systems. Consequently, the time scale of the
oscillation is of the order or larger than the structural relaxation
time scale, therefore the dynamics is completely dissipative.  On the
other hand, in the present case, the structural relaxation time of the
state variable is set by the shear rate and always comparable with the
fluid dynamical time scale, thus the period of the oscillation is
determined by the fluid dynamics.  It is also clear that the stress from
the inertia plays an important role in the thickening phase in the
oscillation.  

\paragraph{Experiments:}

\begin{figure}
\centerline{\includegraphics[width=8cm]{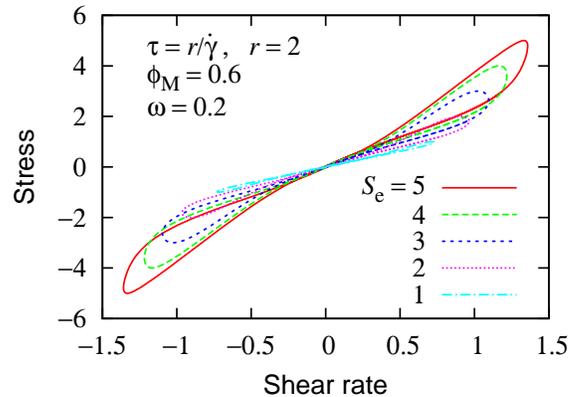}} 
\caption{(Color online)
Hysteresis loops in the shear stress vs the shear rate in
response to oscillatory shear stress for various amplitudes.  The
oscillatory stress $S(t)=S_{\rm e}\sin(\omega t)$ is applied to the
upper plate located at $z=h$ with the lower plate fixed at $z=0$.
The system parameters are $h=0.5$, $\phi_M=0.6$,  $r=2$, and $A=1$.  }
\label{Osci-response}
\end{figure}

As for the hysteresis, Deegan observed the hysteresis loop of the
viscosity under the oscillatory stress for the cornstarch-fluid system
and considered it to be the mechanism for persistent
holes\cite{Deegan-2010}.  His system is thinner and in the regime of
mild shear thickening in comparison with the system studied by Fall et
al\cite{Bonn-2008}.  In the present work, we try to model rather severe
shear thickening in choosing the functional form of the viscosity
Eq.(\ref{eta_phi}), but the present model qualitatively reproduces main
feature of his data, by adjusting the parameters for $\phi_M$ and
$r$ (Fig.\ref{Osci-response} compared with Fig.5 of \cite{Deegan-2010}).
For this set of the parameters, the initial noise decays, thus the
one and two dimensional simulations give the same results.

Regarding the shear thickening oscillation, we could not find any
literature on the experiment which shows clear oscillation.  This may
partly because the system needs to be large enough, typically wider than
$\ell_0$ or a few centimeters, and the inhomogeneity may develops in
three dimensional system, which can obscure the oscillation.  
The noisy fluctuation often observed in rheometer
experiments under constant stress may be explained by this
mechanism\cite{Laun-1991,Lootens-2005}.
Nevertheless, one may easily notice the oscillation around 10 Hz simply
by pouring the dense water-starch mixture out of a container.  We
believe that this oscillation should be explained by the shear
thickening oscillation.  
We are planning experiments
that allow quantitative comparison with our results.
%

Although in the different physical context, the clear oscillatory
flows\cite{Wunenburger-2001} along with a discontinuous transition and
hysteresis\cite{Bonn-1998,Volkova-1999} have been observed in the liquid
crystal system that shows the shear thinning due to the state dependent
viscosity.  Such behavior could be also described using the
phenomenological model like the present one.

Chaotic dynamics has been observed in dilute aqueous solutions of a
surfactant in the experiment under the constant shear rate in the
shear thickening regime\cite{Sood-2000,Sood-2001}.  It was interpreted
as the stick-slip transition between the two states of the fluid
structure, thus physical relevance to the present instability is not
clear, but we also found the chaotic dynamics in the present model in
the case of the constant relaxation time $\tau$ with
the large system width.

\begin{acknowledgements}
This work is supported by KAKENHI(21540418) (H.N.) and the Danish
Council for Independent Research, Natural Sciences (FNU) (N.M.).
\end{acknowledgements}


\end {document}